\documentclass[letterpaper, 10 pt, conference]{ieeeconf}  

\usepackage{graphicx}
\usepackage{adjustbox}
\usepackage{wrapfig}
\usepackage[labelfont=bf,figurename=Fig.,font=small]{caption}
\usepackage{subcaption}
\usepackage{lipsum}
\usepackage{amssymb}
\usepackage{amsmath}
\usepackage{siunitx}
\usepackage{mathtools}
\usepackage{xcolor}
\usepackage{hyperref}
\usepackage[capitalize]{cleveref}
\usepackage[ruled,vlined,linesnumbered]{algorithm2e}
\usepackage[noend]{algpseudocode}
\usepackage{ifthen}
\usepackage{dsfont}
\usepackage[backend=biber,style=numeric-comp,sorting=none]{biblatex}
\usepackage{balance}
\usepackage{dirtytalk}

\usepackage{times}
\usepackage{breqn}

\newcommand{\eqexp}[1]{\underset{\substack{\uparrow\\\mathrlap{\text{\hspace{-1em}#1}}}}{=}}

\bibliography{main}

\IEEEoverridecommandlockouts

\title{\LARGE \bf
Incorporating Data Uncertainty in Object Tracking Algorithms
}

\author{Anish Muthali$^{1}$, Forrest Laine$^{2}$, Claire Tomlin$^{3}$
\thanks{*This work was supported by the DARPA Assured Autonomy Program.}
\thanks{$^{1}$Anish Muthali is a student in the department of Electrical Engineering and Computer Sciences,
        UC Berkeley
        {\tt\small anishmuthali@berkeley.edu}}%
\thanks{$^{2}$Forrest Laine is with the Faculty of Computer Science, Vanderbilt University
        {\tt\small forrest.laine@vanderbilt.edu}}%
\thanks{$^{3}$Claire Tomlin is with the Faculty of Electrical Engineering and Computer Sciences, UC Berkeley
        {\tt\small tomlin@eecs.berkeley.edu}}
}
\begin{document}

\maketitle
\thispagestyle{empty}
\pagestyle{empty}

\begin{abstract}
Methodologies for incorporating the uncertainties characteristic of data-driven object detectors into object tracking algorithms are explored. Object tracking methods rely on measurement error models, typically in the form of measurement noise, false positive rates, and missed detection rates. Each of these quantities, in general, can be dependent on object or measurement location. However, for detections generated from neural-network processed camera inputs, these measurement error statistics are not sufficient to represent the primary source of errors, namely a dissimilarity between run-time sensor input and the training data upon which the detector was trained. To this end, we investigate incorporating data uncertainty into object tracking methods such as to improve the ability to track objects, and particularly those which out-of-distribution w.r.t. training data. The proposed methodologies are validated on an object tracking benchmark as well on experiments with a real autonomous aircraft. 
\end{abstract}

\section{Introduction} \label{sec:intro}

Object tracking can be thought of as the process of inferring the location of objects based on received measurements. This inference problem is naturally posed in a Bayesian framework, but such a setup requires a model about how measurements can be expected to be generated from objects. This is usually modeled in terms of the rate at which false detections are made, the rate at which true detections are missed, and the spatial noise (in measurement space) associated with true detections. Typically all of these statistics are assumed to be the same for all measurements received, although in some cases they can depend on the location of the object which the measurement is associated with, or the location of the measurement itself. One such justification for the former case is that it may be be more likely to miss a detection on objects that are very far away than it is for objects nearby. 

Modeling detections in this way is appropriate for detections coming from sensors such as Radar or Lidar---the errors in measurements are typically due to physical phenomena such as multi-path returns or scattering, which can be related to object or measurement space. However, modern sensors like cameras rely on learned functionality as part of the detection process. This typically takes the form of a Convolutional Neural Network (CNN) mapping the pixel values of the image to 2D boxes which bound detected objects. Considering the CNN as part of the sensor itself allows the resultant bounding boxes to be tracked with standard methods, although we argue that doing so also invalidates the typical measurement model assumed by those tracking methods. A CNN will fail to correctly generate bounding boxes for an image, not necessarily because the objects in that image are near or far, but because there are not similar images represented in the training data. In other words, the measurement model should reflect the fact that learned models may have uncertainty due to the relationship between run-time sensor data and offline training data, not just spatial information like the distance from an object to the sensor. 

In this work, we present our investigation into the hypothesis that accounting for data uncertainty in object tracking methods will result in improved tracking performance, in particular when the task requires tracking objects that are not well-represented in the detector training data. Specifically, we present two object tracking methods which leverage the data uncertainty of CNN-based object detectors for camera inputs: one which is theoretically principled in capturing and leveraging the uncertainty of interest, and another which is a computationally tractable approximation for use in systems such as autonomous vehicles. We evaluate the performance of our proposed framework on a standard benchmark tracking task, as well as by deploying it on an autonomous aircraft tasked with tracking an out-of-distribution object. 
\section{Background and Related Work} \label{sec:related}

The methods we introduce stand on a foundation of existing techniques for a) detecting objects from camera data using CNNs, b) generating estimates of network prediction uncertainty, and c) tracking multiple objects in a Bayesian inference framework. We discuss in this section relevant existing work regarding each of these pillars. 

\subsection{Camera-Based Object Detection}
Effectively all state-of-the art methods for object detection in images rely on deep learning to achieve strong performance. There are a multitude of proposed neural network models and architectures designed for object detection, such as those in \cite{girshick2015fast} \cite{bochkovskiy2020yolov4} \cite{jiang2020sp} \cite{lin2017focal} \cite{postels2019sampling} to cite just a few. In this paper we are primarily interested in using the outputs of these object detectors in a downstream tracking framework, and as such, rely on ``off-the-shelf'' methods for our purposes. In later sections, we detail the particular object detectors used for various implementations of our tracking method. 

\subsection{Uncertainty Characterization in Deep Neural Networks}

Characterization of uncertainty present in deep neural network models is the promise of Bayesian Neural Networks \cite{jospin2020hands}. Producing accurate and computationally efficient estimates of model uncertainty is very much an active problem. We do not claim to solve or contribute towards this goal, but again, leverage existing ``off-the-shelf'' methods for approximating network uncertainty, which we use for downstream tracking tasks. To this end, we primarily investigated using Monte Carlo Dropout to generate estimates of network uncertainty, which has been proposed as a tool for approximating Bayesian Neural Networks \cite{gal2016dropout}. 

Using such a technique, multiple evaluations of a network are made at inference time with dropout enabled, and output variance is estimated using the resultant sampled outputs. The authors in \cite{kendall2017uncertainties} proposed a network architecture for training a network to produce a variance associated with each its output. This method also requires sampling to produce these values of uncertainty. For speed-critical scenarios, sampling-based methodologies may not be suitable, due to the expensive computation required to evaluate the network multiple times.  Sampling-free methods such as that proposed by \cite{postels2019sampling} eliminate the need for repeated sampling. This method injects noise with known variance at the input layer and ``propagates" the variance using gradients such that larger gradients contribute to a higher variance. In our framework, we explore both sampling-based and sample-free approaches to estimating network uncertainty. 

\subsection{Multiple Object Tracking Methods}

The methods for multiple object tracking (MOT) that we derive our approach from can be categorized into two main groups. The first decouples the \emph{measurement association} problem from the track inference problem. In the association phase, the measurements produced by a detector are classified as originating from an object already being tracked, originating from an untracked object, or corresponding to clutter (a false positive). After new tracks are generated if necessary, the inference phase updates object tracks to incorporate the information provided by the newly received measurements, or lack thereof. 
In this class of methods, the measurement association phase can be completed in a greedy manner as demonstrated in \cite{singh2017}, or in a probabilistic manner such as in the PDAF \cite{BARSHALOM1975451} and JPDAF  \cite{barshalom2009} methods. Given measurement assignments, the position update is performed independently for each track. For objects which are assumed to move according to a linear-Gaussian process model, and detections which are assumed to be generated according to a linear-Gaussian measurement model, this process reduces to performing a step of the well-known Kalman filter update each time a detection is associated to a track \cite{kalman1960new}. For more complicated process or measurement models, Extended or Unscented Kalman filter approaches can be used \cite{hoshiya1984structural} \cite{wan2000unscented}. Alternatively, a fixed-lag smoothing approach can be considered, which at each update uses a batch of previous measurements to infer the current state of each object being tracked \cite{dong2011motion} \cite{dellaert2012factor}.

The second class of methods requiring mention are those which reason about the measurement association and track update steps jointly. These approaches model both the number of obstacles and their position as a single random finite set (RFS) \cite{mahler}. The characteristic statistics of the RFS are then updated using the collection of all received measurements, which is also modeled as a RFS. Updating the object RFS is in general not possible in closed form, so tracking algorithms in this class rely on some way of approximately fulfilling the update steps. Examples of such a method include the probability hypothesis density (PHD) filter \cite{mahler} and it's many implementations, such as \cite{clark2006gm} \cite{clark2006convergence}  \cite{ristic2010improved} \cite{particle_methods_overview} \cite{banguvo} \cite{mcmc} \cite{baihmm}. 

An emerging third class of tracking methods includes deep learning based tracking algorithms, which have been shown to perform well on tracking benchmarks \cite{wojke2017simple}. Generally, these deep learning methods may themselves be susceptible to data uncertainty. Therefore we consider the development of these methods as orthogonal to the goal of this paper. 

In all of the existing object tracking methods we are aware of, none are explicitly designed to incorporate the data uncertainty introduced by learning-based object detectors, like we investigate in this work. 
\section{Method} \label{sec:method_a}

The development of our approach for incorporating object detector data uncertainty in a tracking framework is outlined in what follows. We first characterize the approach we chose for training an object detector which can produce uncertainty estimates along with its outputs. We then describe an ideal tracking method which could account for that uncertainty. We show in \cref{sec:results_a} that accounting for network uncertainty significantly improves tracking performance on out-of-distribution (OOD) tracking tasks. 

\subsection{Neural Network Model} \label{subsec:nn}
Our method assumes the existence of neural network trained to generating bounding boxes around detected objects. We require that, in addition to predicted bounding boxes, the network produces an estimate of the uncertainty it has in its detections. This uncertainty could arise from the relationship of the input image to the training data (epistemic uncertainty) or the ambiguity in the detection task on the particular image (aleatoric uncertainty) \cite{kendall2017uncertainties}. Due to its popularity and ease of implementation, we rely on Monte Carlo Dropout methods for producing these estimates of network uncertainty, as discussed in \cref{sec:related}. 

The object detection networks we consider here produce a list of bounding boxes, each corresponding to a location within the input image, as well as a confidence score, valued between $0$ and $1$ \cite{lin2017focal}. Typically a fixed threshold is chosen such that all bounding boxes with confidence score above the threshold are considered true detections. In our proposed Monte Carlo framework, we introduce dropout layers only in the network path responsible for producing bounding box locations. To introduce randomness in which bounding boxes are considered detections or not, we flip the interpretation of the confidence score to instead represent the threshold for considering that output a detection---for each pass of the network, a random variable is sampled from the uniform distribution $U:[0,1]$ for each bounding box output, and if that random variable is less than the detection threshold, that bounding box is considered a detection. If not, the particular bounding box is not included as a detection. 

Running many passes of the network in this way for any particular input image produces a distribution over possible way the network could generate detections for that image, in the form of particles. 

\subsection{ Multiple Object Tracking with Detection Network Uncertainty } 

The objective of a multiple object tracking method is typically to maintain a representation of the distribution 
\begin{equation}
    p(\mathbf{X}_{k}|\mathbf{Z}_{1:k})
\end{equation}
where $\mathbf{X}_{1:k}$ represents the set of all object tracks present at a discrete stage $k$, and $\mathbf{Z}_{1:k}$ is the set of all measurements received between the first stage and stage $k$. This distribution can be expressed, using Bayes rule, as
\begin{equation} \label{eq:bayes_nominal}
\begin{aligned}[b]
    p(\mathbf{X}_{k}|\mathbf{Z}_{1:k}) &= \frac{p(\mathbf{Z}_{k} | \mathbf{Z}_{1:k-1},\mathbf{X}_k)p(\mathbf{X}_k|\mathbf{Z}_{1:k-1})}{p(\mathbf{Z}_{1:k})} \\
    &= \frac{p(\mathbf{Z}_{k} | \mathbf{X}_k)p(\mathbf{X}_k|\mathbf{Z}_{1:k-1})}{p(\mathbf{Z}_{1:k})}.
\end{aligned}
\end{equation}

Here we have used the fact that the measurements at stage $k$ are independent of all previous measurements. The distribution $p(\mathbf{X}_k|\mathbf{Z}_{1:k-1})$ is the predicted distribution of object states at stage $k$, using measurements up to stage $k-1$. This can be expressed as 
\begin{equation} \label{eq:process}
    p(\mathbf{X}_k|\mathbf{Z}_{1:k-1}) = \int_{\mathcal{X}} p(\mathbf{X}_{k-1}|\mathbf{Z}_{1:k-1})p(\mathbf{X}_k | \mathbf{X}_{k-1}) \delta \mathbf{X}_{k-1}.
\end{equation}

Combining these equations, a recursive update can be derived to relate $p(\mathbf{X}_{k-1}|\mathbf{Z}_{1:k-1}) \to p(\mathbf{X}_{k}|\mathbf{Z}_{1:k})$. This update serves as the foundation for all Bayesian object tracking methods.

In the context of tracking bounding box detections arising from a neural network, the bounding boxes produced by the network at stage $k$ would constitute $\mathbf{Z}_k$ in this framework. However, in the framework we are proposing, we are assuming that we no longer work with particular measurements $\mathbf{Z}_k$, but rather a distribution of measurements induced by our network for a particular input. If we let $o_k$ represent the input image to our network at stage $k$, we argue that the distribution we seek to represent is better represented as $p(\mathbf{X}_k|o_{1:k})$. We can represent this distribution as
\begin{equation} \label{eq:good_dist}
    \begin{aligned}[b]
        &p(\mathbf{X}_{k}|o_{1:k}) = \int_{\mathcal{Z}^k} p(\mathbf{X}_k|o_{1:k},\mathbf{Z}_{1:k})p(\mathbf{Z}_{1:k}|o_{1:k}) \delta \mathbf{Z}_{1:k}  \\
        & = \int_{\mathcal{Z}^k} p(\mathbf{X}_k|\mathbf{Z}_{1:k})p(\mathbf{Z}_{1:k}|o_{1:k}) \delta \textbf{Z}_{1:k} \\
        &\eqexp{eq. (2)} \left(\int_{\mathcal{Z}}  \int_{\mathcal{X}} \frac{p(\mathbf{Z}_k|\mathbf{X}_k)p(\mathbf{X}_k | \mathbf{X}_{k-1}) p(\mathbf{Z}_{k}|o_{k})}{p(\mathbf{Z}_{1:k})} \delta \mathbf{X}_{k-1} \delta \mathbf{Z}_k \right) \\
        & \ \ \ \ \ \cdot \Big(\underbrace{\int_{\mathcal{Z}^{k-1}} p(\mathbf{X}_{k-1}|\mathbf{Z}_{1:k-1})p(\mathbf{Z}_{1:k-1}|o_{1:k-1}) \delta \mathbf{Z}_{1:k-1}}_{= p(\mathbf{X}_{k-1}|o_{1:k-1})} \Big)\\
        &= \mathbb{E}_{(\mathbf{X}_{k-1},\mathbf{Z}_{k})\sim P_k} \left[ \frac{p\left(\mathbf{Z}_k|\mathbf{X}_k\right) p\left(\mathbf{X}_k|\mathbf{X}_{k-1}\right)}{p\left(\mathbf{Z}_{1:k}\right)} \right],
    \end{aligned}
\end{equation}
where we define the distribution
\begin{equation}
    P_k = p\left(\mathbf{X}_{k-1} \vert  o_{1:k-1}\right) \cdot p\left(\mathbf{Z}_k \vert o_{k}\right).
\end{equation}
We note that the detections at stage $k$ is conditionally independent of the observations from stages 1 through $k-1$ ($o_{1:k-1}$), given the observation at stage $k$ ($o_k$). The tracks $\mathbf{X}_{1:k}$ are conditionally independent of the observations $o_{1:k}$, given $\mathbf{Z}_{1:k}$.
The form of the resultant distribution in \cref{eq:good_dist} provides a natural framework a particle filter method.

Specifically, we base our approach on the RFS bootstrap filter described by Ristic et al. \cite{particle_methods_overview}. 
At stage $k$, we simply sample $N$ particles from the distribution $P_k$, meaning each particle $i\in\{1,...,N\}$ is associated with a particular instance of the joint obstacle states $\mathbf{X}_{k-1}$ and bounding box outputs $\mathbf{Z}_k$, generated from the network with input image $o_k$. 

Sampling bounding box outputs from the distribution $p\left(\mathbf{Z}_k \vert o_k\right)$ is as simple as evaluating our network according to the process outlined in \cref{subsec:nn}. Other than sampling a different observation $\mathbf{Z}_k$ for each particle, our method is as described in \cite{particle_methods_overview}. Each particle performs its own track birth and track death process, according to the update rules defined by that method. 

\subsection{Track Labeling and Outputting}

An important aspect of the particle filtering approach is to be able to output a consistent ``most-probable'' object representation, which has consistent labels from discrete stage to stage. 

To accomplish this, assume that a most-probable representation of the objects was known at stage $k-1$, and that each of these tracks was labeled with a unique identifier. We refer to this representation as the output frame. The particles that are sampled from the distribution $P_k$, to be used in the measurement update at stage $k$, in general may have more or less objects associated with them than those in the output frame. We assume that the objects contained in any of these particles may or may not correspond to the tracks in the output frame, and if they do, then they share the same label (the procedure for ensuring this is described below). 

After each particle undergoes the birth, death, and measurement update process, and receives a weight, a new output frame needs to be generated. To do this, particles are first binned by the number of tracks they contain. Each bin is associated with a cumulative weight, which corresponds to the total weight of all particles in that bin. The bin with the highest weight is used to generate the new output frame. Some of the tracks of particles in this bin may correspond to previously labeled objects, but some may not. For each object that is labeled, we fit a Gaussian distribution is fit to its location, using the particles in the bin that also contain that object with the same label. A Gaussian Mixture Model is fit to the remaining unlabeled objects appearing in all particles in the bin. We use a number of mixture components equal to the cardinality of the bin (prior to removing labeled objects). If $C$ is the cardinality of this bin, the $C$ highest weighted components are selected for the output frame.  Some of these may correspond to previously labeled objects, and some may correspond to components of the newly fit GMM. The components from the GMM are labeled with new unique identifiers. The mean of the selected components are used in the output frame.

To sample from the new distribution $P_{k+1}$, the cardinality of a new particle is first selected by sampling from a categorical distribution defined by the cumulative bin weights. From the corresponding bin, object locations are sampled from a GMM representing the object locations in that bin. If a particular object in this newly generated particle is close to any labeled object in the output frame, it is assigned the same label, and is left unlabeled otherwise.

\section{Numerical Results} \label{sec:results_a}
We benchmark our uncertainty-enabled tracker using the tools provided by the Waymo Open Dataset 2D Tracking challenge.
\subsection{Neural Network Parameters and Performance}
We trained our bounding box network (RetinaNet with a ResNet-34 backbone) on the entire Waymo Open Dataset training set of 2D images, excluding all images which have pedestrians or cyclists in them. For performance reasons, the images were downsampled to a resolution of $960 \times 640$ instead of the original $1920 \times 1280$. We chose a dropout probability of 0.25 for our network and placed dropout layers after every convolutional layer in the regression component. Our network was trained for 10 epochs and achieved a mean average precision (mAP) score of 0.4582 at 0.5 IoU on the training set.

\subsection{Tracker Parameters}
We define the baseline tracker to be a version of our tracker that does not perform Monte Carlo sampling on the network and does not sample a random variable for detection thresholds. Instead, it uses the same set of detections for each particle and uses a fixed threshold for accepting and rejecting the network's thresholds. We performed hyperparameter search for both trackers. We chose $N = 50$ particles for both trackers as this provided a balanced trade-off between efficiency and performance. We used the metrics described in \cref{subsec:perf} and determined that the optimal (static) confidence threshold for the baseline tracker is 0.7.

\subsection{Tracker Performance}
\label{subsec:perf}
We tested the performance of our trackers on the validation set provided by the Waymo Open Dataset. The goal is to track the location of objects up to stage $k+1$ given camera images for up to stage $k$, repeating this process for $k = 1, \ldots, 200$ in each segment of data. We used Waymo's publicly available tracking metrics computation tools, which computes mean object tracking accuracy (MOTA), mean object tracking precision error (MOTP), fraction of missed objects (Miss), fraction of tracks where the tracking hypothesis changed from previous frames (Mismatch), and fraction of false positives (FP) \cite{bernardin2006multiple}. When testing our tracker, we had to do an extra preprocessing step of relabeling all of the ground truth pedestrians and cyclists as ``vehicles", since our network was only trained to detect vehicles and thus does not have the capability of differentiating between multiple classes. We also separately calculated performance metrics on segments that contain OOD objects (i.e. pedestrians or cyclists).

\subsection{Analysis of Results}
We notice our implementation significantly improves the MOTA, Mismatches, and FP over the baseline. We can hypothesize that our method is likely to be robust to confidence score fluctuations from frame to frame because of the weighting distribution. For example, if a detection has confidence score 0.71 in one timestep but 0.69 in the next, it will no longer be seen by the baseline tracker, but the tracker incorporating data uncertainty would be more robust to small fluctuations such as these. Despite these improvements, we notice a small but non-negligible decrease in performance in MOTP which is likely a consequence of spatial uncertainty.

\begin{figure}
\centering
\begin{subfigure}[b]{0.45\textwidth}
   \includegraphics[width=1\columnwidth]{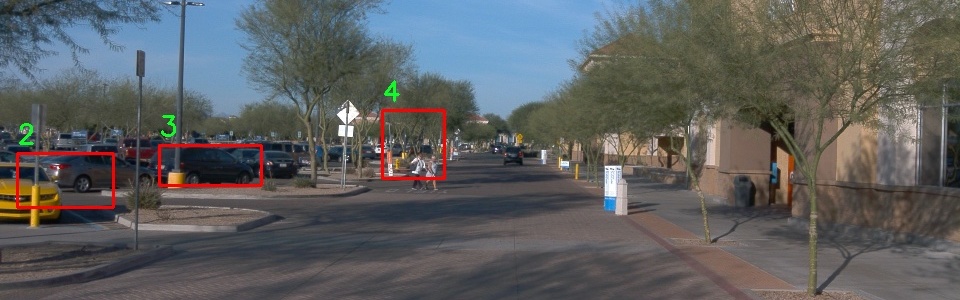}
   \caption{}
   \label{fig:Ng1} 
\end{subfigure}

\begin{subfigure}[b]{0.45\textwidth}
   \includegraphics[width=1\linewidth]{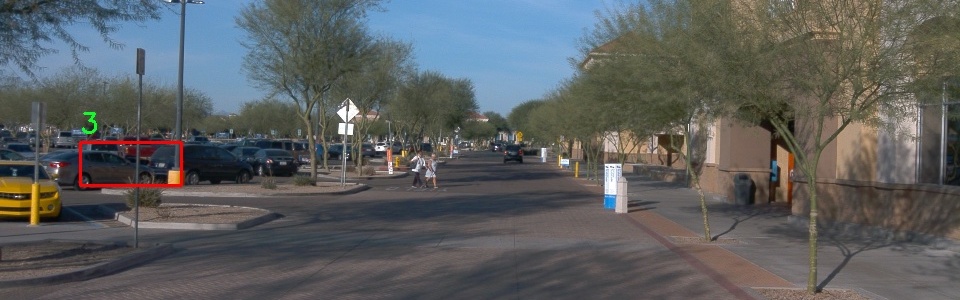}
   \caption{}
   \label{fig:Ng2}
\end{subfigure}

\caption[Ours and theirs]{(a) Representative output frame track locations when tracking detections with uncertainty   \cref{sec:method_a}.
(b) Output frame track locations without accounting for uncertainty.}
\end{figure}

In an example scene depicted in figures \ref{fig:Ng1} and \ref{fig:Ng2}, we can see that, at the same timestep, our tracker is able to pick up the silver and yellow car while the baseline tracker cannot. This is likely a consequence of confidence score fluctuations. Furthermore, our tracker appears to handle OOD objects better, like the two pedestrians in the middle of the road while the network likely cannot detect the pedestrians with high enough confidence consistently to surpass the baseline tracker's threshold. We also notice that the bounding box for track 2 is not definitively encompassing either the yellow or silver car which is likely a consequence of the spatial uncertainty for those two objects.

\begin{table}
\caption{Tracker Performance on All Segments}
\begin{tabular}{ |c|c|c|c|c|c| } 
 \hline
 Tracker & MOTA & MOTP & Miss & Mismatch & FP \\ 
 \hline
 Uncertainty & 0.0853 & 0.5515 & 0.8282 & 0.1654 & 0.2393 \\ 
 \hline
 Baseline & 0.0708 & 0.5516 & 0.8433 & 0.2213 & 0.2923 \\ 
 \hline
 Improvement & \textbf{20.5\%} & \textbf{0.02\%} & \textbf{1.79\%} & \textbf{25.3\%} & \textbf{18.1\%} \\
 \hline
\end{tabular}
\end{table}

\begin{table}
\caption{Tracker Performance on OOD Segments}
\begin{tabular}{ |c|c|c|c|c|c| } 
 \hline
 Tracker & MOTA & MOTP & Miss & Mismatch & FP \\ 
 \hline
 Uncertainty & 0.0741 & 0.5689 & 0.8552 & 0.1657 & 0.2248 \\ 
 \hline
 Baseline & 0.0582 & 0.5582 & 0.8644 & 0.2148 & 0.2786 \\ 
 \hline
 Improvement & \textbf{27.4\%} & -1.93\% & \textbf{1.06\%} & \textbf{22.8\%} & \textbf{19.3\%} \\
 \hline
\end{tabular}
\end{table}

\section{Computationally Efficient Approximation} \label{sec:method_b}

As demonstrated in the previous section, the method presented in \cref{sec:method_a} improved performance on out-of-distribution tracking tasks w.r.t. a baseline implementation not accounting for uncertainty. However, the computational expense of running the algorithm is not tractable for interactive systems, such as autonomous vehicles. To bridge this gap, we developed an alternate method, which retains the conceptual core of the idealized approach, but is amenable to actual deployment. The particular method we develop is tailored for the application of autonomous aircraft taxiing, although many of the design choices made could be generalized to other applications.  

\subsection{Pixel-Level Classification}

A key aspect of the approach presented in \cref{sec:method_a} is the ability to relate object detector data uncertainty with the location of detections in measurement space. While Monte Carlo generation of bounding boxes is one way to accomplish this, it relies on evaluating the neural network many times, which is not feasible for real-time systems. Therefore we turned to the sample-free variance propagation technique \cite{postels2019sampling} to generate estimates of uncertainty in a single pass of the network. This method approximates the variance of a univariate Gaussian distribution centered at each output of the network, representing the network uncertainty for that value. 

For bounding box outputs, however, we found Gaussian approximations of each bounding box location to be a poor representation of the distribution of possible network outputs.  To address this, we propose training a pixel-level occupancy classifier, which is tasked with classifying whether each pixel in the input image corresponds to empty space or an obstacle. Variance propagation can be used to generate a 2D Gaussian distribution over the penultimate layer of the classifier, which can be propagated through the final soft-max layer to generate a value which can be interpreted as the probability of that particular pixel being occupied. 

Applying this computation to the output of every pixel generates a probability map of obstacle occupancy over the entire image. For any given input pixel with resultant probability of occupancy $p$, we convert that value $p$ into a binary occupancy variable $o \in \{ 0,1\}$, where $o=1$ (representing occupied) if $p\geq0.5$, else $o=0$, and an uncertainty measure $u$, where $u=2(1-p)$ if $o=1$, and $u=2p$ if $o=0$. In this way, the uncertainty value $u$ is close to $1$ when $p\approx0.5$, and $u$ is close to $0$ when $p\approx0$ or $p\approx1$. This change of variable makes it easier to reason about these outputs in the object tracking procedure.

\subsection{Tracking Occupancy Grids}

This transformed output can be interpreted as a binary mask, indicating the pixel locations of obstacles, along with uncertainty estimates, where lower uncertainty corresponds to higher confidence in the binary classification. This form of the network output, in theory, contains information regarding the spatial distribution of false negatives and false positives due to data uncertainty. For a particular image, if the output of the detector corresponding to a particular region of the image is classified as non-occupied, but has high uncertainty values, this region may correspond to a missed detection. Conversely, a region classified as occupied with a high uncertainty may correspond to a false positive. 

The output at each pixel can be treated as an independent measurement on potential objects being tracked, and standard measurement association and measurement update methods can be applied. In particular, each pixel corresponding to a detection can be assigned to either an existing track or new track, and pixels corresponding to empty space can be assigned to all nearby tracks. 

As in \cref{sec:method_a}, we assume that objects are tracked in the image plane. We assume that the uncertainty in object tracks locations are maintained as a Gaussian distribution, and the uncertainty in object tracks is maintained as a probability of existence. We assume that objects evolve according to a linear-Gaussian process model. Under these modeling assumptions, we propose the following particle-filter based method for efficiently updating these object track representations using pixel level measurements.

Assume that a representation of an object track is known at some discrete time stage $k$. To update this representation to reflect measurements received at stage $k+1$, the Gaussian representation of the object state is propagated through the linear-Gaussian process model, producing a Gaussian distribution over the predicted object state at stage $k+1$. Using a particle approach, many particles representing object track state vectors are sampled from this distribution. Each particle is then weighted by the product of the probabilities of measuring the independent measurements, conditioned on the object track state represented by the particle. 

\begin{figure}
    \centering
    \includegraphics[width=\columnwidth]{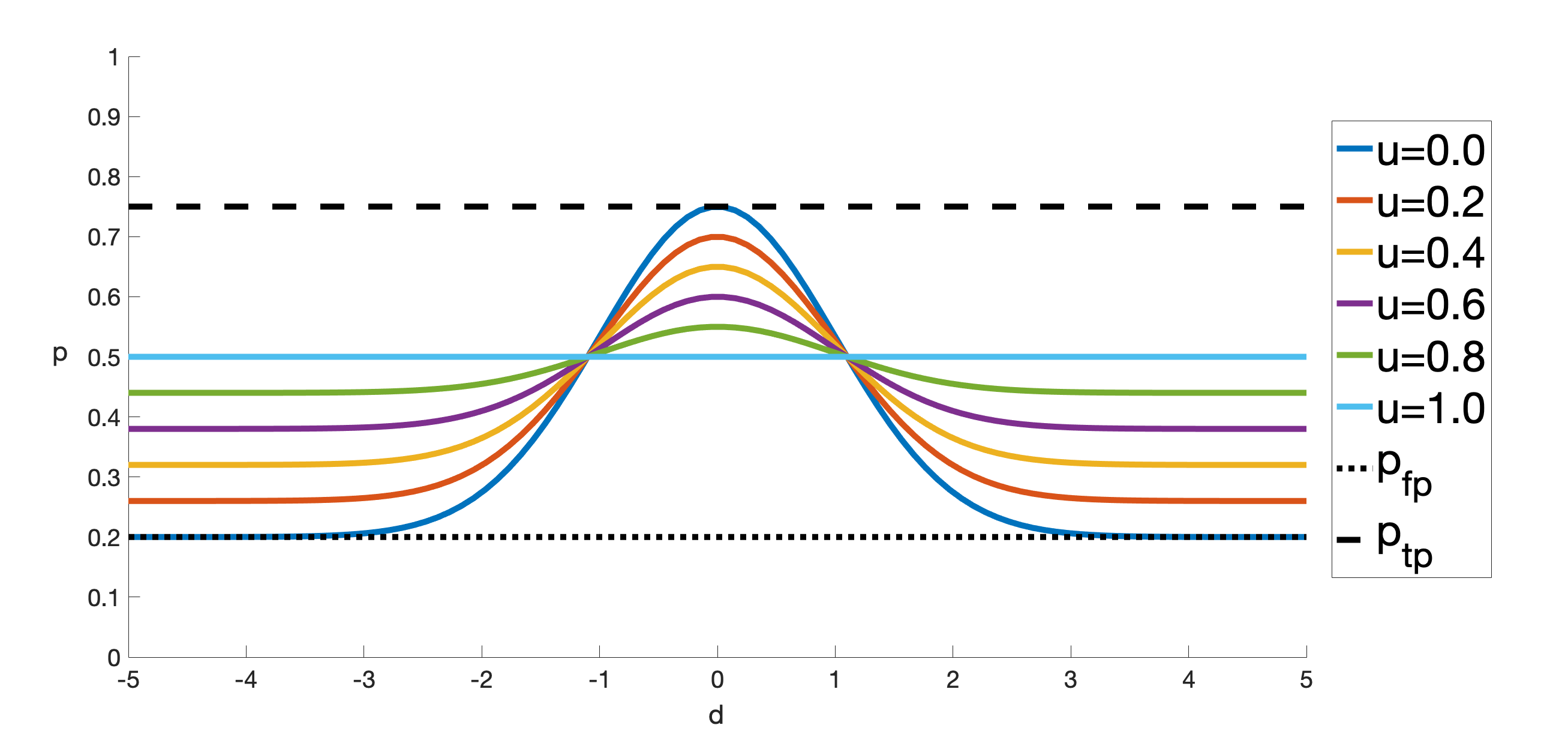}
    \caption{The measurement likelihood function $p(o=1,u|x)$ as a function of the set distance $d$, for various uncertainty values. Here $p_{fp} = 0.2$, $p_{tp}=0.75$, and $\sigma=1$. When $u=0$, the likelihood peaks at $p_{tp}$ when $d=0$, and decays to $p_{fp}$ as $|d|$ increases. When $u\to 1$, the likelihood function converges towards $p(o=1,u|x)=0.5$ for all $d$.}
    \label{fig:meas}
\end{figure}

We define conditional measurement likelihood for an individual pixel $i$ with output $o_i,u_i$, conditioned on a particular object track state $x$ as the following:
\begin{equation} \label{eq:measurement_model}
\begin{aligned}
    p(o_i=1,u_i | x) &= \\
    p_{fp} + &(p_{tp}-p_{fp}) e^{-\frac{d_i^2}{2\sigma^2}} (1-u_i) + \frac{u_i}{2} \\
    p(o_i=0,u_i | x) &= 1 - p(o_i=1,u_i | x),
\end{aligned}
\end{equation}
where $d_i$ is defined to be the set distance between the object track bounding box and the 2D region of the image plane corresponding to the pixel $i$, and $p_{tp}$ is the nominal probability of a true positive detection, $p_{fp}$ is the nominal probability of a false positive, and $\sigma$ is a shaping parameter. A visualization and rationalization of this measurement likelihood function is given in \cref{fig:meas}. 

Specifically, the particle weighting $w$ for a particle $x$ is defined as 
\begin{equation} \label{eq:weighting}
    w = \prod_{i=1}^N \ p(o_i,u_i|x),
\end{equation}
where $N$ is the number of pixels associated with the object track represented by particle $x$. Particles which overlap with many pixels with $o_i=1$, and don't overlap with many pixels with $o_i=0$ will be weighted highly. Pixels with very high uncertainty will effectively assign all particles the same weight. After all particles are weighted, a posterior distribution of the track distribution is generated by fitting a Gaussian distribution to the weighted particles. 

To assist in the measurement association process, we first cluster occupied pixels with neighboring (or nearly neighboring) occupied pixels. We additionally include in each cluster any unoccupied pixels which surround the cluster. We allow the unoccupied pixels to be part of multiple clusters, but occupied pixels must belong to exactly one cluster. These clusters are then either associated to existing tracks, or used to birth new tracks. For the association process, we use a greedy approach, although others could be considered. The product in \cref{eq:weighting} is then taken over all pixels belonging to the associated cluster.  

The final step of the tracking procedure is to update the probability of existence for all tracks. The probability of existence is maintained as the ratio of probability that an object exists over probability that an object does not exist. This ratio is updated according to the following update:

\begin{equation}
    LR_{k+1} = LR_{k} \cdot \frac{\bar{w}}{\prod_{i=1}^N p(o_i,u_i|\emptyset)}.
\end{equation}
Here, $\bar{w}$ is the average weight taken over all particles, and \begin{equation}
\begin{aligned}
    p(o_i=1,u_i|\emptyset) &= p_{fp}(1-u_i) + \frac{u_i}{2} \\
    p(o_i=0,u_i|\emptyset) &= 1-p(o_i=1,u_i|\emptyset),
\end{aligned}
\end{equation} 
is defined to be the probability of measuring pixel $i$ with occupancy $o_i$ and uncertainty $u_i$ when there is no nearby object. In practice, when the likelihood ratio $LR_{k+1} << 1$ after a measurement update, the object track under consideration is killed. 
\section{Implementation on Autonomous Aircraft} \label{sec:results_b}
To validate that the method proposed in \cref{sec:method_b} is suitable for deployment on real autonomous systems, we deployed an implementation on an autonomous Cessna Grand Caravan 208B developed by Boeing. This integration was performed as a demonstration for the DARPA Assured Autonomy program. 

The demonstration aimed to test the capability of our approach in tracking out-of-distribution objects in an interactive, online setting. Similar to the setup in \cref{sec:results_a}, we trained an object detector based on the architecture presented in \cite{postels2019sampling} to produce pixel level detections on a data set provided by Boeing, which consisted of ground vehicles on airport runways. This data set comprised of  images labeled in the form of bounding boxes. In order to produce per-pixel labels, we converted the bounding boxes to a binary mask, where pixels in the interior of labeled bounding boxes were labeled as occupied, and all other pixels were labeled as unoccupied. 

The training data held out all instances of other aircraft, so that at test time, these objects were considered out-of-distribution. The particular task posed to the object tracking system was to track an aircraft that entered the field of view, and follow that aircraft down an airport runway. Video footage of our implementation successfully accomplishing this task is included in the supplementary video submission. Our method was able to detect and track the aircraft, despite the detector never being trained on that type of object. Furthermore, both the object detection network and object tracking algorithm were able to run at a rate faster than 5hz, which was suitable for use in the downstream control task.

\begin{figure}
    \centering
    \includegraphics[width=\columnwidth]{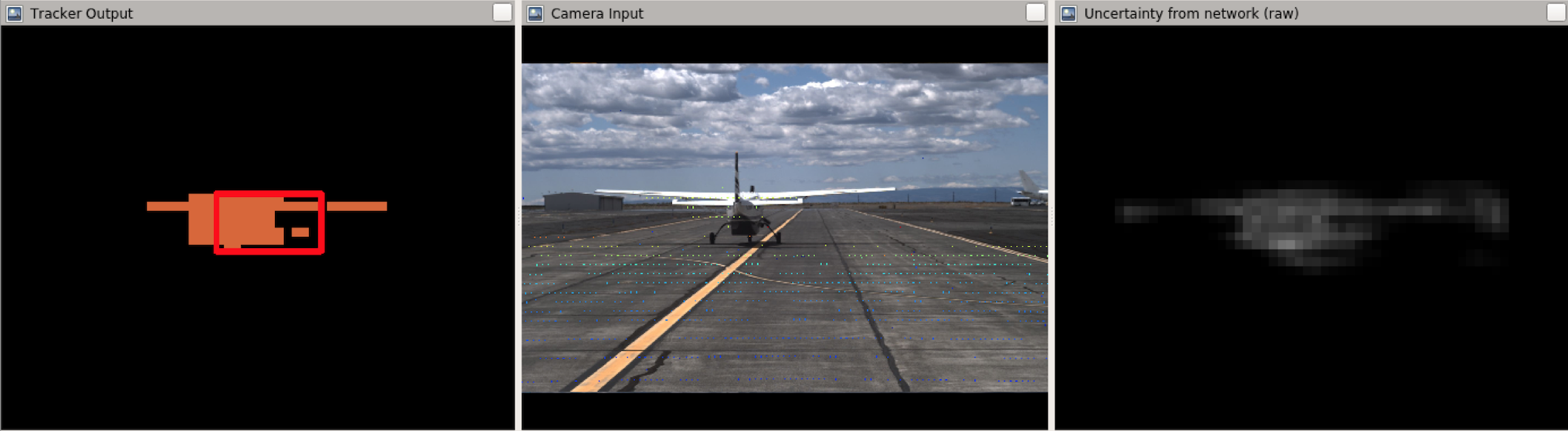}
    \caption{Visualization of our method running on the Boeing platform. The central window displays the RGB input fed into the neural-network based object detector. The window on the left displays the binary mask indicating which pixels are detected to be occupied by an object (in this case another aircraft). Overlaid on top of the detections is the mean estimate of the object track bounding box. The window on the right visualizes the per-pixel uncertainty map associated with the network outputs.}
    \label{fig:screengrab}
\end{figure}
\section{Conclusion} \label{sec:conclusion}
We presented a claim that incorporating data uncertainty from learning based object detectors in an object tracking framework can improve tracking performance, especially when tracking objects not well represented in the training data of the object detector. We supported this claim with results on standard tracking benchmark. We then proposed an approximate but tractable method, and implemented it on an autonomous aircraft. 

\section*{Acknowledgments}
We thank Jim Paunicka, Isaac Chang, Michael McGivern, Matt Moser, Dragos Margineantu, Jordan Stringfield, Ashley Abril, Doug Stuart, and and the entire team at Boeing for all their help and contributions in demonstrating our method on the Caravan Test Platform. 

\printbibliography

\end{document}